\documentclass[conference]{IEEEtran}
\IEEEoverridecommandlockouts
\usepackage{cite}
\usepackage{amsmath,amssymb,amsfonts}
\usepackage{graphicx}
\usepackage{textcomp}
\usepackage{xcolor}
\usepackage{graphicx}
\usepackage{amsmath}
\usepackage[noend]{algpseudocode}
\usepackage{algorithmicx,algorithm}
\usepackage{amsthm}
\usepackage{amsfonts}
\usepackage{subfigure} 
\usepackage{color}
\usepackage{cite}
\usepackage{setspace}
\newtheorem{Lemma}{Lemma}
\newtheorem{Theorem}{Theorem}
\usepackage{amssymb}
\usepackage{color}
\graphicspath{{images/}}
\usepackage{booktabs}
\usepackage{multirow} 
\usepackage{makecell}
\usepackage{threeparttable}
\usepackage[numbers,sort&compress]{natbib}
\newtheorem{remark}{Remark}
\usepackage{stfloats}
\usepackage{amsmath}
\usepackage{amssymb}
\usepackage{enumitem}
\def\BibTeX{{\rm B\kern-.05em{\sc i\kern-.025em b}\kern-.08em
    T\kern-.1667em\lower.7ex\hbox{E}\kern-.125emX}}
\begin{document}

\title{\huge Outage Probability Analysis of Uplink Heterogeneous Non-terrestrial Networks: A Novel Stochastic Geometry Model}

\author{Wen-Yu Dong, Shaoshi Yang*,~\IEEEmembership{Senior Member,~IEEE}, Wei Lin, Wei Zhao, Jia-Xing Gui,\\
 and Sheng Chen,~\IEEEmembership{Life Fellow,~IEEE}	%
 	\thanks{This work was supported in part by the Beijing Municipal Natural Science Foundation (No. L242013), and in part by the Open Project Program of the Key Laboratory of Mathematics and Information Networks, Ministry of Education, China (No. KF202301). \textit{(Corresponding author: Shaoshi Yang}}
	\thanks{W.-Y. Dong, S. Yang, W. Lin, W. Zhao, and J.-X Gui are with the School of Information and Communication Engineering, Beijing University of Posts and Telecommunications, with the Key Laboratory of Universal Wireless Communications, Ministry of Education, and also with the Key Laboratory of Mathematics and Information Networks, Beijing 100876, China (E-mails: wenyu.dong@bupt.edu.cn, shaoshi.yang@bupt.edu.cn, linwei@bupt.edu.cn, wei.zhao@bupt.edu.cn, jiaxing.gui@bupt.edu.cn).} %
	\thanks{S. Chen is with the School of Electronics and Computer Science, University of Southampton, Southampton SO17 1BJ, U.K. (E-mail: sqc@ecs.soton.ac.uk).} %
	\vspace*{-5mm}

}
\maketitle

\begin{abstract}
 In  harsh environments such as mountainous terrain, dense vegetation areas, or urban landscapes, a single type of  unmanned aerial vehicles (UAVs) may encounter challenges like flight restrictions, difficulty in task execution, or increased risk. Therefore, employing multiple types of UAVs, along with satellite assistance, to collaborate becomes essential in such scenarios. In this context, we present a stochastic geometry based approach for modeling the heterogeneous non-terrestrial networks (NTNs) by using the classical   binomial point process and introducing a novel point process, called Mat{\'e}rn hard-core cluster process (MHCCP). Our MHCCP possesses both the exclusivity and the clustering properties, thus it can better model the aircraft group composed of multiple clusters. Then, we derive closed-form expressions of the outage probability (OP) for the uplink (aerial-to-satellite) of heterogeneous NTNs. Unlike existing studies, our analysis relies on a more advanced system configuration, where the integration of beamforming and frequency division multiple access, and the shadowed-Rician (SR) fading model for interference power, are considered. The accuracy of our theoretical derivation is confirmed by Monte Carlo simulations. Our research offers fundamental insights into the system-level performance optimization of NTNs.
\end{abstract}


\section{Introduction}
Non-terrestrial networks (NTNs), encompassing unmanned aerial vehicles (UAVs), high-altitude platforms (HAPs), and satellite networks, are commonly utilized for a variety of purposes such as remote sensing, navigation, disaster management, and other commercial applications \cite{9861699}. However, in specific environments, singular-type low-altitude UAVs may confront challenges such as flight restrictions, task execution difficulties, or heightened risks. For instance, in rugged terrains like mountainous regions, densely vegetated areas, or urban landscapes, one UAV type may outperform others, necessitating the integration of multiple low-altitude UAV types in a heterogeneous configuration. Therefore, employing multiple types of low-altitude UAVs in a heterogeneous configuration is crucial.

However, in challenging terrains, substantial shadow fading occurs due to obstructions like buildings, trees, and mountains, rendering conventional fading models such as Rayleigh, Rician, and Nakagami less applicable. In heterogeneous non-ground networks, the presence of multiple types of UAVs, their spatial distribution, and communication characteristics all contribute to the complexity of analysis. Therefore, conventional research methods often rely on discrete event simulations or network simulators, which lack the use of mathematical methods \cite{chen2010uav, ma2013simulation} to comprehensively study the potential features of networks. Additionally, these methods tend to impose notable constraints when applied to specific network contexts. Recent studies have demonstrated that stochastic geometry \cite{1995Stochastic} and random geometric graphs \cite{2002Random} have emerged as highly effective tools for addressing the aforementioned challenges.

In contrast to Rayleigh and Nakagami-m fading, shadowed-Rician (SR) fading \cite{1623307} has been proven to be more appropriate for the statistical characterization of satellite channels. This model has been found to be applicable across several frequency bands, such as S-, L-, Ku-, and Ka-band, making it a versatile choice for modeling satellite communication channels.  While the SR model is extensively employed in satellite link research, system-level analysis is understudied. With LEO satellites located at different altitudes, the work \cite{2021Stochastic} investigated the joint coverage probability from satellites to satellite gateways in remote areas and then to anchor base stations, given that the satellite-to-gateway link is subject to SR fading. The authors of \cite{9678973} derived the downlink OP of the LEO satellite communication system under SR fading based on a BPP distribution and optimized the system throughput under visibility and outage constraints. Nevertheless, similar to majority of the works utilizing SR models, these two papers solely examined SNR while neglecting the consideration of interference. The works \cite{7869087} and \cite{8068989} approximated the squared SR model with a Gamma random variable and solves the interruption probability using a Gamma function model. Song \emph{et al.} \cite{song2022cooperative} employed the Nakagami-m fading model rather than the SR fading model for the purpose of estimating statistical values of interference. 
Hence, it is evident that the interference analysis with the SR model remains unexplored. 

Motivated by the aforementioned discoveries, this study investigates the heterogeneous NTN system, comprising two distinct groups of airplanes and one satellite. To better model such a heterogeneous NTN, we introduce a novel point model called Mat{\'e}rn hard-core cluster process (MHCCP), which is derived by combining or integrating type-\uppercase\expandafter{\romannumeral2} Mat{\'e}rn hard-core point process (MHCPP)  and Mat{\'e}rn cluster process (MCP). The location distributions of these two aircraft groups are described as the MHCCP for one group and the BPP for the other group. Furthermore, beamforming and frequency division multiple access (FDMA) are employed to improve network efficiency. Since in this heterogeneous NTN system, nodes operating within the same subchannel share the same frequency band, we explicitly investigate the effects of multi-user interference (MUI) under SR fading channels. Our contributions can be succinctly summarized through the following four key points. 

\begin{itemize}
\item Unlike existing studies, we consider a more advanced and comprehensive system model, which involves many factors, such as network deployment, multi-access mechanisms, beam models, and channel models, when analyzing the performance of heterogeneous NTNs.
\item Different from what is often overlooked in previous research work,  we consider interference in the presence of SR fading, while conducting the uplink performance analysis for heterogeneous satellite-aerial networks (SANs). 

\item 
Compared to other similar articles, we introduce a novel point process model, referred to as the MHCCP, that offers a comprehensive representation of both the clustering and exclusion properties of nodes. 

\item We verify through Monte Carlo simulation that our analysis of uplink performance for heterogeneous SANs is highly accurate, and discuss the impact of different parameters.
\end{itemize}	

The remainder of this paper is organized as follows: In Section~\ref{S2}, the system model is presented. Section~\ref{S3} provides our primary performance analysis results. In Section~\ref{S4}, we provide numerical results to verify our theoretical derivations and to study the effect of key system parameters. Finally, our conclusions are drawn in Section~\ref{S5}.
\section{System Model}\label{S2}
\begin{figure}[tbp]
	
	\centerline{\includegraphics[width=0.87\linewidth]{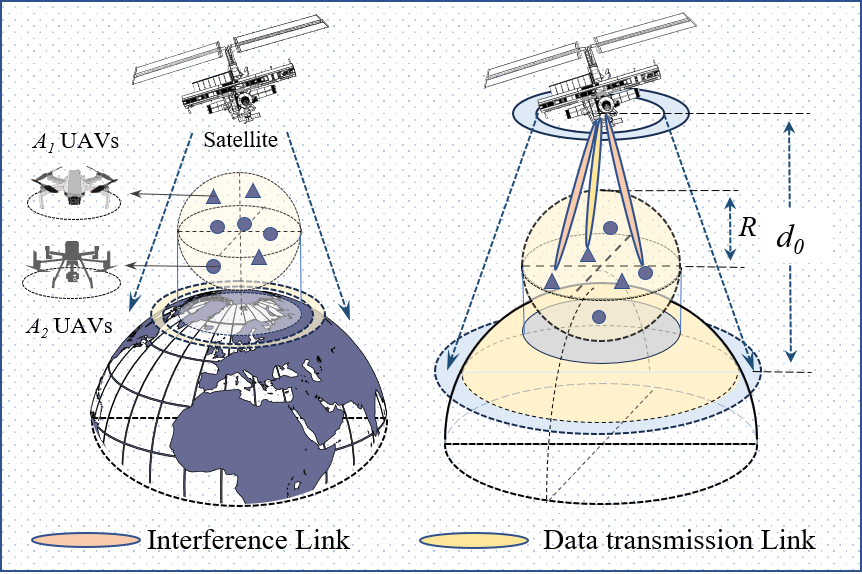}}
	\vspace*{-2mm}
	\caption{Illustration of the heterogeneous NTN system.}
	\label{fig:1}
	\vspace*{-5mm}
\end{figure}

We consider a SAN composed of a satellite (LEO satellite \emph{S}) as well as two heterogeneous UAV groups (\emph{A$_1$} and \emph{A$_2$}), as shown in Fig.~\ref{fig:1}. The two groups of heterogeneous UAVs exhibit distinct features in terms of their parameters and transmitter capacity. For instance, in tasks like geological disaster detection, UAV group \emph{A$_1$} operate in clustered formations to survey surface areas, while UAV \emph{A$_2$} functions individually, offering flexibility to explore challenging terrains such as canyons, mountainous regions, and indoor structures for localized monitoring and reconnaissance. Considering that UAV coverage areas are much smaller than the distances to satellites, we opt to model both \emph{A$_1$} and \emph{A$_2$} within the same spherical space. This approach simplifies the description without affecting the derived results, even when two heterogeneous UAV swarms are not situated in the same spherical space. The trajectories of satellites can be predicted and the wide angle of their receiver antenna is large. Our analysis focuses on the uplink connection between UAV groups and a specific satellite.

\begin{figure}[t]
	\centerline{\includegraphics[width=0.87\linewidth]{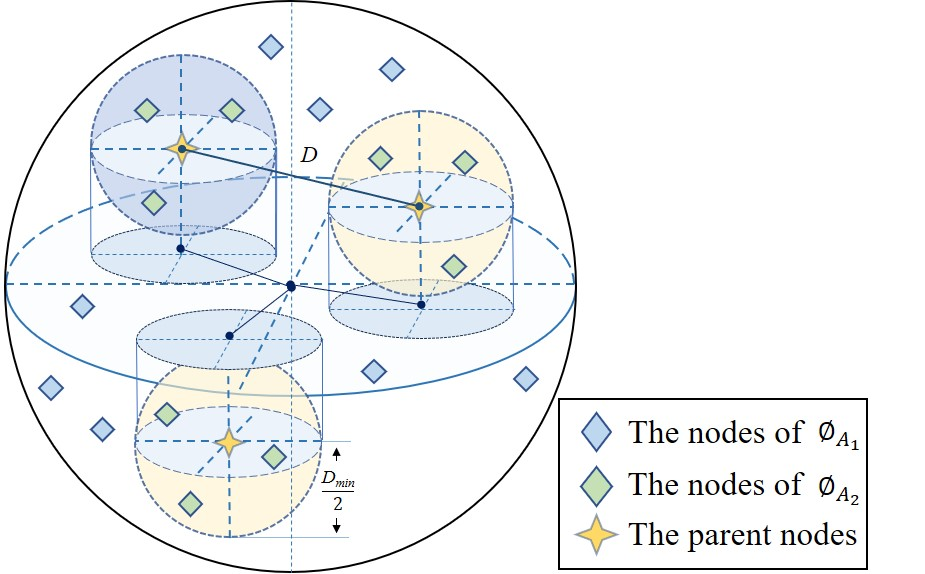}}
		\vspace*{-2mm}
	\caption{Illustration of node positions in heterogeneous aircraft network.}
	\label{fig:2}
	\vspace*{-5mm}
\end{figure}

\subsection{Topology Model}\label{S2.1}

\subsubsection{Deployment of $A_1$ and $A_2$}  

According to \cite{srinivasa2009distance}, it is impossible to accurately describe networks with a finite number of nodes and limited area using the PPP. In practical applications, $A_1$ typically consists of a finite number of nodes and operates within a limited-size region. Therefore, we posit that $A_1$ adheres to the principle of the BPP $\phi_{A_1}$, with a numerical value of $N_1$, within the spherical region $\mathcal{V}$ of radius $R$. Furthermore, the MHCCP is utilized in order to represent the clustered networking of $A_2$, which allows for the maintenance of a secure distance between different clusters within $A_2$, facilitating cooperative operations. Fig.~\ref{fig:2} illustrates the distribution of node locations for $A_1$ and $A_2$.

Next, we outline the three-step approach involved in the construction of MHCCP, denoted as $\phi_{A_2}$.

In the first step, candidate points are generated in a uniform manner using a homogeneous PPP with the density $\lambda_1$, within the spherical space $\mathcal{V}$. The radius of $\mathcal{V}$ is $R_1$, and the volume of $\mathcal{V}$ is $V_1$. In relation to the number of candidate points, denoted as $N_c$, the probability mass function of the Poisson distribution can be represented as $\mathbb{P}(N_{c}=s)=\frac{\lambda_{1}V_1}{s!}\mathrm{exp}{(-\lambda_{1}V_1)}$, where $\emph{s}$ is the number of the candidate points.

In the second step, each candidate point is assigned with an independent mark, randomly selected from a uniform distribution within the range $[0, ~ 1]$. Subsequently, we exclusively retain the point with the smallest mark within a confined spherical space with the radius $D_{\min}$ by eliminating the others. Specifically, for a specific point $Q$ serving as the center of a small spherical repulsion space with the radius $D_{\min}$, if there are other candidate points within this spherical repulsion space, their mark values are compared, and only the point with the smallest mark is preserved within each spherical repulsion space. This process iterates for each candidate point until each spherical repulsion space contains only one point. After this removal process, the distance $D$ between any two points is larger than or equal to $D_{\mathrm{min}}$, and we have constructed the MHCPP, denoted as $\phi_c$, with the density $\lambda_2$ which can be expressed mathematically as $\lambda _{2}=\frac{1-\mathrm{exp}(-\frac{4}{3}\pi D_{\mathrm{min}}^{3}\lambda _1)}{\frac{4}{3}\pi D_{\mathrm{min}}^{3}}$.

In the final step, we exploit the principle of MCP by considering the points in \emph{$\phi_c$} as the parent points, and uniformly generate subpoints within the sphere of radius $D_{\min}/2$ around each parent point. The number of subpoints in each cluster follows a Poisson distribution with parameter $\overline{c}$. Through this clustering process, we complete the MHCCP construction, and all the subpoints generated form $\phi_{A_2}$. The relationship between the density of the parent points, $\lambda_2$, and the density of $\phi_{A_2}$, $\lambda_3$, is expressed by $\lambda_3=\lambda_2\overline{c}$.

\subsubsection{Multi-access Mechanism} 

Multi-access mechanisms in aerial-to-satellite communications play an important role in improving the uplink transmission performance. To enhance the network's connectivity and capacity, the FDMA is employed, implementing frequency reuse across a total of $K$ orthogonal frequency channels ($K \le N_1$). To enable communication with the satellite on a cluster basis, the number of frequency bands in A2 is an integer multiple of the clusters, i.e., $N_2/K$ is expected to be an integer which is multiple of $\overline{c}$. For the purpose of simplifying calculations, we assume that $N_1/K + N_2/K$ aerial transmitters share a common frequency channel, thereby causing co-channel interference. Hence, we randomly select one channel from the $(N_1+N_2)/K$ available channels for investigation, and denote the sets of nodes from the two UAV groups sharing the same frequency band as $\phi_1$ and $\phi_2$, respectively.

\subsection{Channel Model}\label{S2.2}

\subsubsection{Directional Beamforming Modeling} 

To enhance the strength of the received signal and optimize transmission performance, directional beamforming is employed at aerial transmitters and satellite receivers. To simplify processing, the actual antenna is approximated using a sector antenna model. Due to the extensive coverage capability of the satellite, it is reasonable to assume that the satellite receiver can effectively capture the signals transmitted by the UAVs through the primary lobe. The overall directional gain of link $l$ is denoted as $D_l$, and $\theta$ represents the angular breadth of the primary lobe of transmitter. In this context, $G_t$ and $g_t$ refer to the transmitter array gains of the main and side lobes, respectively. Similarly, $G_r$ denotes the receiver array main lobe gain.

The value of $D_m$ for the link from a given target node $m$ to the satellite receiver is equal to the product of $G_t$ and $G_r$, i.e., $D_m=G_t G_r$. On the other hand, the value of $D_{x_l}$ for any other interference node $x_l$ is determined by the directivity gains of the main and side lobes of the antenna beam pattern. Accordingly, we have the probability distribution model expression of $D_{x_l}$ as:
\begin{align}\label{eqDG} 
	D_{x_l} =& \left\{ \begin{array}{cl}
		G_t G_r ,	& {\rm P}_{M,M} = \frac{\theta}{2\pi} , \\ 
		g_t G_r	, & {\rm P}_{S,M} = 1 - \frac{\theta}{2\pi} ,
	\end{array} \right.
\end{align}
where the indices $M$ and $S$ denote the main and side lobes, respectively, and ${\rm P}_{t,M}$, $t\! \in\! \{ M,S\}$, denotes the probability of the link in state $\{t,M\}$.

\subsubsection{Channel Fading} 

In this paper, the presence of abundant shadows due to complex terrain, along with the predominant propagation path being the Line-of-Sight (LoS) path, closely aligns with the characteristics of SR fading. Therefore, utilizing SR fading for modeling the aerial-satellite link in complex terrain at low altitudes is reasonable. We define $2 c$ as the mean power of the multi-path component excluding the LoS component, $\Omega$ as the average power of the LoS component, and $q$ as the Nakagami-m fading parameter. The small-scale fading is represented as $|h|^{2}$ in this study. The probability density function (PDF) of $|h|^{2}$ can be mathematically stated as \cite{zhang2019performance}
\begin{eqnarray}\label{eqSF} 
	f_{|h|^2}(x) = \kappa\, \exp(-\beta x)\, {_1F_1}(q; 1; \delta x),
\end{eqnarray}
where  $ \kappa =\frac{(2 c q)^q}{2 c(2 c q+\Omega)^q}$, $ \delta=\frac{\Omega}{2 c(2 c q+\Omega)}$, $\beta=\frac{1}{2 c}$ and ${_1F_1}\left(\cdot;\cdot;\cdot\right)$ is the confluent hypergeometric function of the first kind..

\subsection{SINR Model}	\label{S2.3}

In accordance with the aforementioned model, the SINR, a critical metric for evaluating wireless link performance, can be expressed as:
\begin{eqnarray}\label{eqSINR} 
	\mathrm{SINR} = \frac{p_m D_m\left|h_{m}\right|^2 d_{m}^{-\alpha}} {I+\sigma^2},
\end{eqnarray}
where $I=\sum_{l\in \{1,2\}}\sum_{x_l\in\phi _l {\backslash} \{m\}} p_{x_l}D_{x_l}\left|h_{x_l }\right|^2 d_{x_l}^{-\alpha}$, $l$ is the index of a particular UAV gruop in  $A_1$ and $A_2$, $m$ and $x_l$ are the target node and interference nodes, $p_m$ and $p_{x_l}$ are the transmit power at $m$ and $x_l$, while $d_{m}$ is the distance between $m$ and the satellite, $d_{x_l}$ is the distance between $x_l$ and the satellite, $\alpha$ is the path-loss exponent, and $\sigma^2$ is the strength of additive white Gaussian noise (AWGN).

\section{Performance Analysis}\label{S3}

This section analyzes the OP  for the proposed heterogeneous SAN model. Without loss of generality, we assume uniform transmission power level for all transmitters within each aircraft group, except for the target transmitter. Specifically, except for the target transmitter $m$ whose power is set to $p_m$, the power levels for aircraft in the two groups, $\phi_{A_1}$ and $\phi_{A_2}$, are given by $p_1$ and $p_2$, respectively.

The OP refers to the likelihood that the SINR at the receiver is insufficient to meet the minimum SINR required for successful data transmission. Thus, the OP of the aerial-to-satellite link is defined by\begin{eqnarray}\label{eqOP} 
	P_{\mathrm{out}} \triangleq \mathbb{P}(\mathrm{SINR}\le T) = \mathbb{P}\left( \frac{p_m D_m\left|h_{m}\right|^2 d_{m}^{-\alpha}} {I+\sigma^2} \le T \right) , 
\end{eqnarray}
where $T$ is the SINR threshold. The following theorem derives a useful expression for OP.

\begin{Theorem}\label{T1}
	The outage probability for an arbitrarily located aerial node under the SR fading channel is given by
	\begin{align}\label{eqT1}
		P_{\mathrm{out}} \triangleq & \,\, \mathbb{P}(\mathrm{SINR}\le T) \nonumber \\
		= & \sum\limits_{k=0}^{\infty} \frac{\Psi (k)}{(\beta -\delta )^{k+1}} \Gamma (k+1) \sum\limits_{t=0}^{k+1}\binom{k+1}{t} (-1)^t \nonumber \\
		&	\times \mathbb{E}\left[\exp\big(-s(I + \sigma^2)\big)\right] ,		
	\end{align}
	where $\Psi (k)=\frac{(-1)^{k} \kappa \delta^{k}}{(k!)^{2}}{\rm Ps}(1-q)_{k}$, $s=\frac{t \zeta (\beta -\delta) T d_{m}^{\alpha}}{p_m D_m}$ and $\zeta = (\Gamma (k+2))^{-\frac{1}{k+1}}$.
\end{Theorem}

\begin{proof}
	Using the Kummer's transform of the hypergeometric function, we can rewrite the PDF of $|h|^{2}$ as $f_{|h|^{2}}(x) = \sum\limits_{k=0}^{\infty}\Psi (k) x^{k} \exp( -(\beta - \delta) x$, where $\Psi (k) = \frac{(-1)^{k}\kappa \delta^{k}}{(k!)^{2}} {\rm Ps}(1-q)_{k}$ and  the Pochhammer symbol is defined as ${\rm Ps}(x)_{n}=\Gamma(x+n)/\Gamma(x)$. Then, the CDF of $|h|^{2}$ can be represented as\begin{align}\label{eqA1} 
		F_{|h|^{2}}(x) =& \sum_{k=0}^{\infty} \Psi (k) \int\limits_{0}^{x} t^{k} \exp(-(\beta - \delta ) t) \mathrm{d}t \nonumber \\
		=& \sum\limits_{k=0}^{\infty} \frac{\Psi (k)}{(\beta -\delta )^{k+1}} \gamma (k+1,(\beta - \delta )x) .
	\end{align}

	Therefore, we can obtain $P_{\rm out}$ as
	\begin{align}\label{eqA2} 
		& P_{\mathrm{out}} \triangleq \mathbb{P}\left( \frac{p_m D_m |h_{m}|^2 d_{m}^{-\alpha}}{I + \sigma^2} \le T   \right) \nonumber \\
		&= \mathbb{E}\left[\kappa \sum\limits_{k=0}^{\infty} \frac{\Psi (k)}{(\beta -\delta )^{k+1}} \gamma \left(k+1,(\beta -\delta ) \frac{T (I + \sigma^2) d_{m}^{\alpha}}{p_m D_m}\right)\right] \nonumber \\  
		& \overset{\mathrm{(a)}}{\approx} \mathbb{E}\Bigg[ \sum\limits_{k=0}^{\infty} \frac{\Psi (k)}{(\beta -\delta )^{k+1}} \Gamma (k+1) \nonumber \\ 	
		&	\times \left(1 - \exp\left(-\frac{\zeta (\beta - \delta) T (I + \sigma^2) d_{m}^{\alpha}}{p_m D_m} \right)\right)^{k+1}\Bigg]  \nonumber \\
		& \overset{\mathrm{(b)}}{=} \sum\limits_{k=0}^{\infty} \frac{\Psi (k)}{(\beta - \delta)^{k+1}} \Gamma (k+1) \sum\limits_{t=0}^{k+1} \binom{k+1}{t}(-1)^t \nonumber \\
		&	\times \mathbb{E}\left[\exp(-s(I + \sigma^2))\right] , 
	\end{align}
	where (a) is approximated by using $\gamma (k+1,x) < \Gamma (k+1)(1-\exp(-\zeta x))^{k+1}$ \cite{alzer1997some}, $\zeta=(\Gamma(k+2))^{-\frac{1}{k+1}}$, and (b) is obtained from the binomial theorem with $s=\frac{t \zeta (\beta -\delta ) T d_{m}^{\alpha}}{p_m D_m}$. This completes the proof.
\end{proof}

Given that the distances between aerial nodes in finite-area networks are typically limited to a few kilometers, while the distance to the satellite extends to the range of hundreds or even thousands of kilometers, it is reasonable to assert that the latter is significantly greater in magnitude than the former. Therefore, it is assumed that all the aerial transmitters possess an equal transmission distance to the satellite, i.e., $d_m=d_{x_l}=d_0$. Subsequently, we obtain
\begin{align}\label{eqT1s2} 
	\mathbb{E}\left[\exp(-s(I + \sigma^2))\right] 
	=& \exp(-s \sigma^2) \mathcal{L}_I(s) ,
\end{align}
where $\mathcal{L}_I(s)$ is the Laplace transform of the cumulative interference power $I$ that is expressed in Lemma~\ref{L1}.

\begin{Lemma}\label{L1}
	The Laplace transform of random variable $I$ is:
	\begin{align}\label{eqLT} 
		\mathcal{L}_I(s) =& \mathbb{E}[\exp(-sI)] \nonumber \\
		=& \left(M_2(1)\right)^{n_1} \exp\big(\lambda_{3} V_1 (M_2(2) - 1)\big) ,
	\end{align}
	where $	V_1=\frac{4\pi R_1^3}{3}$, $\lambda_{3} = \overline{c}\,\, \frac{1 -\exp\big(-\frac{4}{3} \pi D_{\mathrm{min}}^{3} \lambda_1\big)}{\frac{4}{3}\pi D_{\mathrm{min}}^{3}}$, $M_2(l) = M_1(\mu_l )\frac{\theta}{2\pi} + M_1(\nu_l)(1 - \frac{\theta}{2\pi}), \, l\in\{1,2\}$, with $	M_1(t_l) = \frac{(2 c q)^q (1 + 2 c t_l)^{q-1}}{\big((2 c q + \Omega )(1 + 2 c t_l)-\Omega\big)^q} , \,t_l\in \left\{ \mu_l,\nu_l\right\}$
	, $\mu_l=-sp_ld_0^{-\alpha}G_t G_r$ and $\nu_l=-sp_ld_0^{-\alpha}g_t G_r$,	and when the target UAV node is in $\phi_1$ and  $\phi_2$, $n_1$ is given by $\frac{N_1}{K}-1$ and $\frac{N_1}{K}$, respectively.
\end{Lemma}

\begin{proof}
	\begin{align}\label{eqB1} 
		& \mathcal{L}_I(s) = \mathbb{E}\left[\exp(-s I)\right] \nonumber \\
		& \hspace*{3mm} = \mathbb{E}\!\! \left[\! \exp\! \bigg(\! -s\!\! \sum\limits_{l\in \{1,2\}} \sum\limits_{x_l\in\phi_l{\backslash} \{m\}}\!\! p_{l} D_{x_l} \left|h_{x_l}\right|^2 d_{x_l}^{-\alpha} \bigg)\! \right]\!\! .
	\end{align}

	As the point process and the fading process are independent of each other, $\mathcal{L}_I(s)$ can be expressed as
	\begin{align}\label{eqB2} 
		& \mathcal{L}_I(s) = \mathbb{E}\left[\!\prod_{l\in \{1,2\}} \prod_{x_l\in\phi _l \!{\backslash}\! \{m\}}\!\! \!\!\!\!\! \mathbb{E}_{|h_{x_l}|^2}\!\!\left[\mathrm{exp}\!\left(\!-s p_{l} D_{x_l}\!\left|h_{x_l }\right|^2 \! d_{x_l}^{-\alpha}\right)\right]\!\right] \nonumber \\
		& \hspace*{1mm} \overset{\mathrm{(a)}}=\! \mathbb{E}_{N_l}\!\!\left[\prod_{l\in \{1,2\}}\prod_{x_l\in\phi _l \! {\backslash}\! \{m\}}\!\!\!\! \mathbb{E}_{D_{x_l},|h_{x_l}|^2} \!\!\left[\mathrm{exp}\left(-t_l\left|h_{x_l }\right|^2 \right)\right] \!\right]\!\! , \!
	\end{align}
	where (a) is obtained by assuming that all the aerial transmitters have the same transmission distance and denoting $t_l=s p_{l} D_{x_l}d_{0}^{-\alpha}$.
	
	As shown in \cite{2003A}, the moment-generating function of the SR model is defined as $M_S(x)=\mathbb{E}\left[\exp(-x S)\right]=	\frac{(2 c q)^q(1+2 c x)^{q-1}}{((2 c q+\Omega )(1+2 c x)-\Omega )^q}$. Thus, we further obtain
	\begin{align}
		\mathcal{L}_I(s)\! =& \mathbb{E}_{N_l}\! \Bigg[\! \prod_{l\in \{1,2\}}\! \prod_{x_l\in\phi _l {\backslash} \{m\}} \!\!\! \mathbb{E}_{D_{x_l}}\!\! \bigg[\! \underset{M_1(t_l)}{\underbrace{\frac{(2 c q)^q(1+2 c t_l)^{q-1}}{((2 c q\! +\! \Omega )(1\! +\! 2 c t_l)\! -\!\Omega )^q}}} \! \bigg] \! \Bigg] \nonumber \\
		\! \overset{\mathrm{(b)}}=& \mathbb{E}_{N_l}\! \Bigg[\! \prod_{l\in \{1,2\}}\! \prod_{x_l\in\phi _l {\backslash} \{m\}} \!\! \bigg[ \! \underset{M_2(l)}{\underbrace{M_1(\mu_l)\frac{\theta}{2\pi}\! +\! M_1(\nu_l)\Big(1\!-\!\frac{\theta}{2\pi}\Big)}}\! \bigg]\! \Bigg] \nonumber\\
		\! =& \mathbb{E}_{N_1}\! \bigg[\prod_{x_1\in\phi_1 {\backslash} \{m\}}\! M_2(1)\bigg] \mathbb{E}_{N_2}\bigg[\prod_{x_2\in\phi_2 {\backslash} \{m\}} M_2(2) \bigg] \nonumber \\
			\end{align}
			
	\begin{align}\label{eqB3} 
		\! \overset{\mathrm{(c)}}=& \sum_{n=1}^{n_1} \binom{n_1}{n} P_I^n (1-P_I)^{n_1-n} \left(M_2(1)\right)^{n}\nonumber \\
		\! & \times \sum_{n_2=0}^{\infty} \frac{(\lambda_{3}V_1)^{n_2}}{n_2!}\mathrm{exp}(-\lambda_3V_1)(M_2(2))^{n_2} \nonumber \\
		\! \overset{\mathrm{(d)}}=& \left(M_2(1)\right)^{n_1}\exp\big(\lambda_{3}V_1 (M_2(2)-1)\big),
	\end{align}
	where (b) is obtained by denoting $\mu_l=s p_l d_0^{-\alpha} G_t G_r$ and $\nu_l=s p_l d_0^{-\alpha}g_t G_r$, (c) is obtained by the fact that $\phi_{A_1}$ follows the BPP with $P_I$ being the proportion of the interfering nodes to the total number of nodes in $\phi_{A_1}$, $\phi_{A_2}$ follows the PPP with the density of $\lambda_{3}$ and the volume of the spherical space $\mathcal{V}$ is $V_1=\frac{4\pi R_1^3}{3}$, and (d) is obtained by the fact that all the nodes in $\phi_{A_1}$ are interference nodes and hence $P_I=1$. This completes the proof.
\end{proof}

\begin{remark}\label{R1}
The MHCCP model exhibits strong repulsion between clusters, which is determined by the minimum distance separating them. Consequently, the system's total number of nodes cannot exhibit unlimited growth as $\lambda_1$ rises. In order to ascertain the upper limit of nodes in this system, we differentiate the function $\lambda_3$ over $\lambda_1$, yielding  $\frac{\mathrm{d}{\lambda_3}}{\mathrm{d}\lambda_1} = \overline{c}\big(\frac{4}{3}\pi D_{\mathrm{min}}^{3}\big)^2\exp\big(-\frac{4}{3}\pi D_{\mathrm{min}}^{3}\lambda_1\big)$. Notably, $\frac{\mathrm{d}{\lambda_3}}{\mathrm{d}\lambda_1}\ge 0$. Moreover, $\lim\limits_{\lambda_1\rightarrow\infty} \frac{\mathrm{d}{\lambda_3}}{\mathrm{d}\lambda_1} =0$. Hence, the upper limit of $\lambda_3$ can be established as $\lim\limits_{\lambda_1\rightarrow\infty}\lambda_3 = \frac{3\overline{c}}{4\pi D_{\mathrm{min}}^3}$, whose validity will be further examined in the simulation discussed in Section~\ref{S4}.
\end{remark}

By substituting (\ref{eqT1s2}) and (\ref{eqLT}) into (\ref{eqT1}) and noting the definitions of Lemma~\ref{L1}, we can obtain the analytical closed-form expression for the OP.

\section{Numerical Results}\label{S4}

In this section, we validate the derived theoretical expressions by Monte-Carlo simulations with a total of 50,000 iterations. Unless otherwise explicitly specified, the default simulation system parameters detailed in Table~\ref{Table2} are used. The outcomes obtained from the analytical expressions derived in Section~\ref{S3} are labeled as `Analysis', whilst the Monte Carlo findings are labeled as `Simulation'.

\begin{table}[!h]\scriptsize
		\vspace*{-4mm}
	\caption{Default Parameter of Simulation System.} 
	\label{Table2} 
	\vspace*{-3mm}
	\begin{center}
	\begin{tabular}{|c|c|c|}
		\hline
		\textbf{Notation} & \textbf{Parameter} & \textbf{Values} \\
		\hline
		$d_0$	& Distance between UAVs and satellite & 300km \\
		\hline
		$R_1$ & Radius of $\mathcal{V}$ & 10km \\
		\hline
		$D_{\mathrm{min}}$ & Minimum distance of candidate pairs & 1km \\
		\hline
		$p_1$, $p_2$ & Power of transmitters in $A_1$ and $A_2$ & 20dBW, 19dBW \\
		\hline
		$\lambda_{1}$ & Density of candidate points & $10^{-11}$ \\
		\hline
		$\mathcal{SR}(c,q,\Omega)$ & SR fading model & $\mathcal{SR}(0.158,1,0.1)$ \\
		\hline
		$\alpha$ & Path-loss exponent & 2 \\
		\hline
		$T$ & SINR threshold & -18dB \\
		\hline
		$\sigma^2$ & AWGN’s power spectral density & -160dBm/Hz \\
		\hline
	\end{tabular}
	\label{tab1}
		\vspace*{-5mm}
\end{center}

\end{table}
\begin{figure}[!t]
	\begin{center}
		\includegraphics[width=0.9\columnwidth]{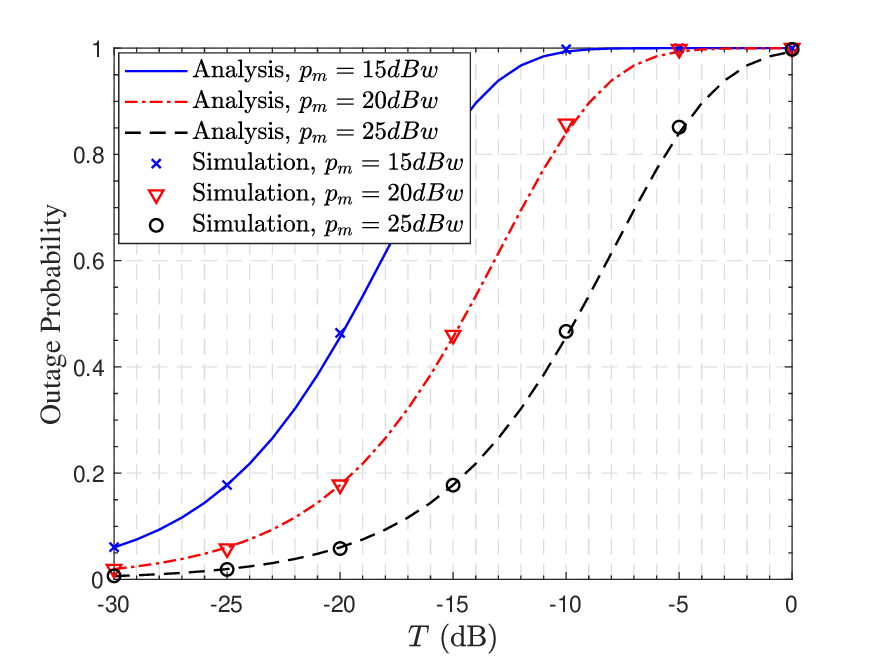}
	\end{center}
	\vspace*{-5mm}
	\caption{Outage probability as function of SINR threshold $T$, given different transmit power $p_m$ of target node.}
	\label{fig:3}
	\vspace*{-7mm}

\end{figure}

\begin{figure}[!b]
	\vspace*{-7mm}
	\begin{center}
		\includegraphics[width=0.9\columnwidth]{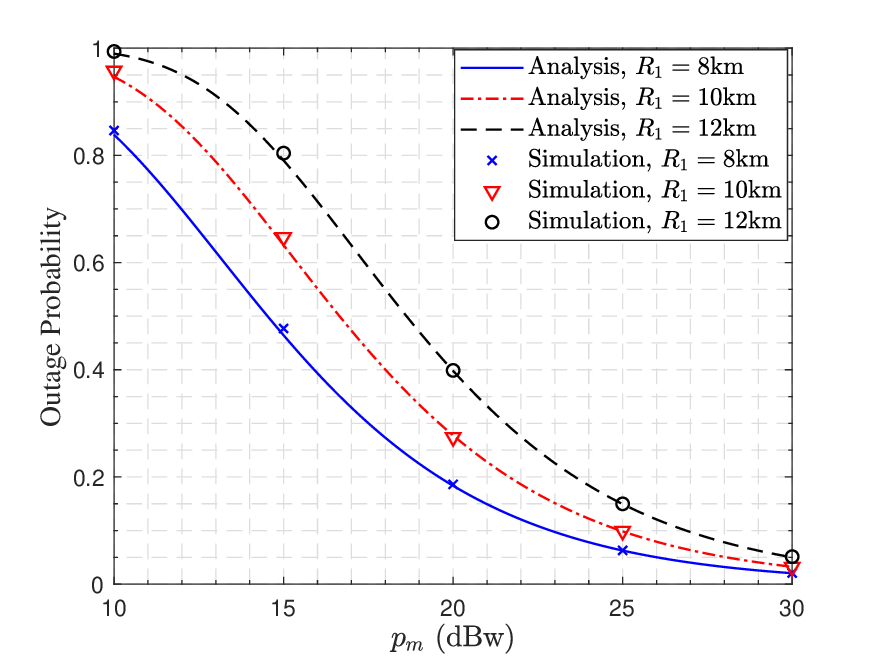}
	\end{center}
	\vspace*{-5mm}
	\caption{Outage probability as function of target node transmit power $p_m$, given different values of spherical space's radius $R_1$.}
	\label{fig:4}
	\vspace*{-1mm}
\end{figure}
These simulations are utilized to generate the plots depicting the simulated OP performance of the heterogeneous aerial-to-satellite uplink.  We observe that the analytical results exhibit a high degree of concordance with the corresponding simulation results, hence bolstering the credibility and soundness of our theoretical investigation presented in Section~\ref{S3}.

\begin{figure}[!t]
	\vspace*{-1mm}
	\begin{center}
		\includegraphics[width=0.9\columnwidth]{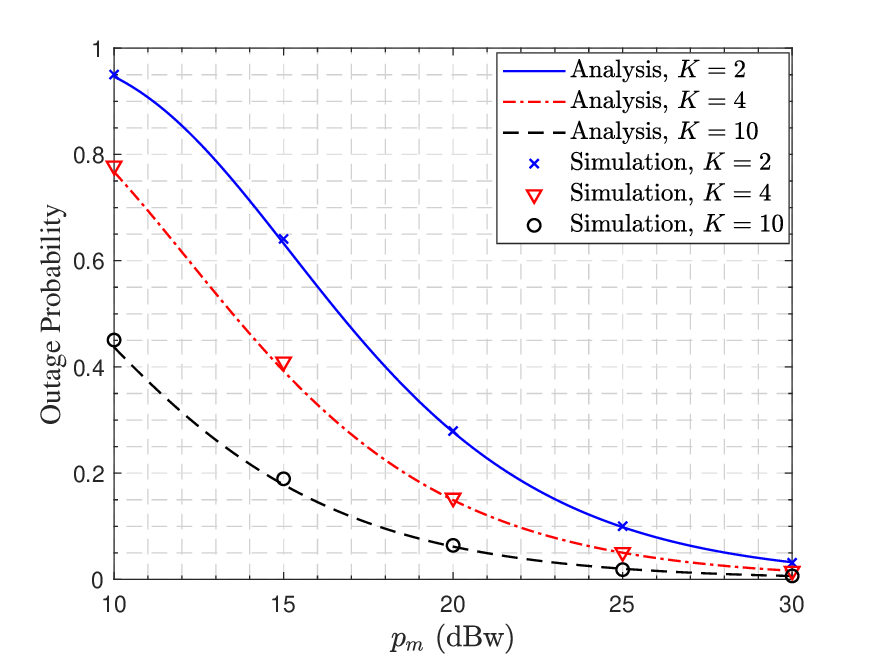}
	\end{center}
	\vspace*{-5mm}
	\caption{Outage probability as function of target node transmit power $p_m$, given different numbers of frequency channels $K$.}
	\label{fig:7}
	\vspace*{-4mm}
\end{figure}

First Fig.~\ref{fig:3} depicts the OP as the function of the SINR threshold $T$, given three different values of the target node's transmit power $p_m$. It can be seen from Fig.~\ref{fig:3} that as the SINR threshold $T$ increases, the OP also increases. This is due to the inverse relationship between $T$ and the possibility of achieving a SINR that surpasses the given threshold value. Hence, once the threshold value reaches to a specific level, the likelihood of disruption occurring in the link between the aerial transmitter and the satellite reaches its maximum value of 1. Furthermore, increasing the power of the target transmitter leads to an increase in its SINR and this reduces the risk of communication interruption. Therefore, with an increase in $p_m$, the OP curve exhibits a rightward shift.  

Next Fig.~\ref{fig:4} plots the OP as the function of $p_m$, given three different values for the radius $R_1$ of the spherical space $\mathcal{V}$. As expected, increasing $p_m$ decreases the OP. Also it can be seen that the expansion of the distribution space of aerial transmitters leads to a noticeably worse OP performance. This is because the increased availability of space for the MHCCP deployment of air nodes results in a higher number of transmitters concurrently attempting to access the satellite, leading to a higher MUI and consequently a worsen OP performance.
\begin{figure}[!t]
	\vspace*{-1mm}
	\begin{center}
		\includegraphics[width=0.9\columnwidth]{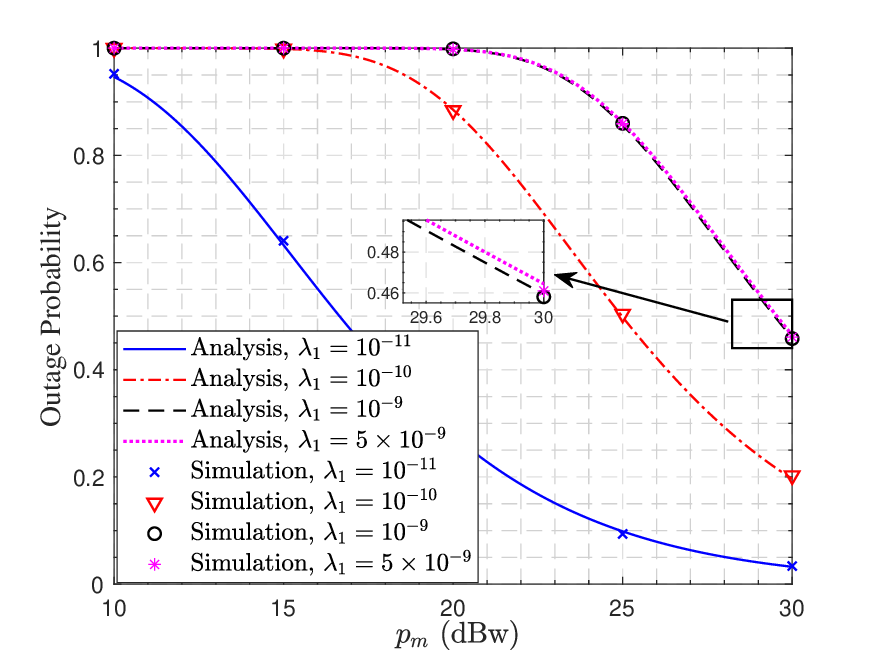}
	\end{center}
	\vspace*{-5mm}
	\caption{Outage probability as function of target node transmit power $p_m$, given different numbers of frequency channels $K$.}
	\label{fig:8}
	\vspace*{-5mm}
\end{figure}

Then fig.~\ref{fig:7}~provides a clear visual representation of the influence of the number of frequency channels $K$ on the OP, indicating that increasing $K$ decreases the OP. Evidently, providing more frequency channels enables more transmitters operate in orthogonal access mode, thereby resulting in a reduction of the interference towards the intended transmitter. Consequently, the SINR of the target link increases, diminishing the likelihood of interruption. 

Fig.~\ref{fig:8} depicts the OP as the function of $p_m$, given four different values for the density of candidate points $\lambda_{1}$. As expected, increasing $\lambda_{1}$ increases number of nodes, which results in greater MUI and consequently increases the OP. However, as can be seen from Fig.~\ref{fig:8}, when $\lambda_{1}$ increases beyond $10^{-9}$, the OP appears saturated. This is because the imposed minimum spacing $D_{\mathrm{min}}$ prevents the number of nodes from infinitely escalating as the density increases.

\section{Conclusions}\label{S5}

In this paper, we have proposed a tractable approach for analyzing the outage probability  for the uplink of heterogeneous satellite-aerial networks. 
Our proposed novel point model can better describe the clustering distribution of aircraft Our novel contribution has been two-fold. Then, the interference analysis template we provide for the SR model is applicable to many other related derivation studies. Finally, the exact closed form solution expression for the interruption probability we derived can be directly applied in similar scenarios, thereby quickly obtaining performance analysis of the network.  Our study therefore offers theoretical guidance and valuable insights for the actual deployment of heterogeneous satellite-aerial networks.

\small
\bibliographystyle{IEEEtran}
\bibliography{IEEEabrv, reference}
\end{document}